**Gate Tunable Relativistic Mass and Berry's phase in Topological Insulator Nanoribbon Field Effect Devices**

Luis A. Jauregui[1,2], Michael T. Pettes[3,†], Leonid P. Rokhinson[4,1], Li Shi[3,5], Yong P. Chen[1,4,2,*]

[1] Birck Nanotechnology Center, Purdue University, West Lafayette, IN 47907

[2] School of Electrical and Computer Engineering, Purdue University, West Lafayette, IN 47907

[3] Department of Mechanical Engineering, University of Texas at Austin, Austin, TX 78712

[4] Department of Physics, Purdue University, West Lafayette, IN 47907

[5] Materials Science and Engineering Program, University of Texas at Austin, Austin, TX 78712

[†] Current address: Department of Mechanical Engineering, University of Connecticut, Storrs, CT 06269

[*] To whom correspondence should be addressed: yongchen@purdue.edu

**Abstract**

Transport due to spin-helical massless Dirac fermion surface state is of paramount importance to realize various new physical phenomena in topological insulators, ranging from quantum anomalous Hall effect to Majorana fermions. However, one of the most important hallmarks of topological surface states, the Dirac linear band dispersion, has been difficult to reveal directly in transport measurements. Here we report experiments on $Bi_2Te_3$ nanoribbon ambipolar field effect devices on high-k $SrTiO_3$ substrates, where we achieve a gate-tuned bulk metal-insulator transition and the topological transport regime with substantial surface state conduction. In this regime, we report two unambiguous transport evidences for gate-tunable Dirac fermions through $\pi$ Berry's phase in Shubnikov-de Haas oscillations and effective mass proportional to the Fermi momentum, indicating linear energy-momentum dispersion. We also measure a gate-tunable weak anti-localization (WAL) with 2 coherent conduction channels (indicating 2 decoupled surfaces) near the charge neutrality point, and a transition to weak localization (indicating a collapse of the Berry's phase) when the Fermi energy approaches the bulk conduction band. The gate-tunable Dirac fermion topological surface states pave the way towards a variety of topological electronic devices.

*Introduction*

The extraordinary electronic properties of topological insulators[1-3] (TIs) make them a unique class of materials relevant for applications such as low power electronic devices, spintronics[1] and fault-tolerant quantum computation[4,5]. TIs feature topologically non-trivial surface states, where carriers are massless relativistic particles with linear energy-momentum band dispersion and with spins locked perpendicular to their momentum[2,3]. Existence of such helical Dirac fermion surface states have been experimentally confirmed by angle-resolved photoemission spectroscopy (ARPES)[6,7] and scanning-tunneling microscopy[8]. Transport measurements in bulk crystals of TI (such as $Bi_2Te_3$)[9] have met more challenges to probe the surface states, because of the non-insulating bulk conduction. In order to reduce the bulk contribution and enhance the surface contribution in TI transport, a number of efforts have been made by growing bulk single-crystals with fine-tuned composition[10-13], adding compensating dopants[14,15], growing ultrathin films and nanostructures such as nanowires[16,17] or nanoribbons (NRs)[11,18], and using electrical gating[19-21] to tune the Fermi energy ($E_F$). Nanostructures, with a high surface-to-volume ratio, have been shown to enhance the surface conduction contribution. However, in previous reports, unambiguous measurements of the topological surface states (e.g $\pi$-Berry's phase and Dirac fermion dispersion) remain challenging. In this work, we combine nanostructures with ultrahigh-$\kappa$ gate dielectric by placing $Bi_2Te_3$ NRs on $SrTiO_3$ (STO) substrates. We realize the topological transport regime where bulk is insulating and surface substantially contributes to the conduction, and report clear evidence of the Dirac fermion nature of the gate-tunable topological surface states, directly revealing the characteristic linear energy-momentum dispersion via density-dependent effective mass measurements. The excellent gate tunability of $E_F$ (from surface states to bulk conduction band) has also enabled us to observe a predicted transition from the usual weak anti-localization (WAL) behavior to weak localization (WL), reflecting a change in the pseudospin texture and Berry's phase of the quantum coherent charge carriers [22,23].

The $Bi_2Te_3$ NRs were synthesized by catalyst-free vapor solid method, following an approach similar to that of Kong et al.[24] for thin films. NRs are grown out of the plane of the substrate and they are individually transferred to other substrates or support for further studies, using an electrochemically sharpened tungsten probe and an optical microscope. We have used transmission electron microscopy (TEM) to determine the structural characteristics of the $Bi_2Te_3$ NRs. Low-magnification TEM reveals a typical width of ~100–230 nm and a length exceeding 10μm (Fig. 1a). High-resolution TEM confirms the single-crystallinity, with an atomic spacing of 2.2 Å (Fig. 1b, marked with arrows) along the $[11\bar{2}0]$ direction, similar to previously grown single-crystal $Bi_2Te_3$ NRs[21]. The Fourier transform (FFT) of the high-resolution TEM image reveals a high-quality single-crystalline structure with hexagonal symmetry (inset of Fig. 1b). We have also performed Raman spectroscopy with a 532nm excitation laser on $Bi_2Te_3$

NRs (placed on SiO$_2$/Si substrates). Figure 1c shows a representative spectrum. We observe three Raman peaks at 62 cm$^{-1}$, 104 cm$^{-1}$ and 137 cm$^{-1}$, in excellent agreement with the Raman optical phonon modes ($A_{1g}^1$, $E_g^2$ and $A_{1g}^2$) previously measured in bulk crystals[25] and exfoliated thin flakes[26,27] of Bi$_2$Te$_3$, further confirming the good crystalline quality of our NRs. Fig. 1d shows the atomic force microscope (AFM) image of a typical NR with thickness ~ 30 nm and width ~130 nm. Most of the electronic transport data presented below (unless otherwise noted) are measured from this representative Bi$_2$Te$_3$ NR, which is transferred onto a 500-μm-thick STO substrate (with very high relative dielectric permittivity κ at low temperatures[28]) and fabricated into a back gated field effect device (device #1). Similar results have also been obtained in several other devices measured.

*Temperature dependence and field effect.* Effective gate control of topological surface states is highly desired for investigations of novel quantum transport and many device applications of TIs [4,21]. The optical image of device #1 and a schematic of its cross section are shown in the lower and upper insets of Fig. 2a, respectively. Figure 2a also shows the ambipolar field effect measured in the NR at 2K, where the decreasing backgate voltage ($V_g$) tunes the carriers from n-type to p-type, with the resistance (R) peaking at the charge neutrality point ($V_{CNP}$ ~ -17V). The typical gated R modulation ratio is ~ 3-10 in our field effect devices. For $V_g$ = 0V, the temperature (T) dependence of the resistance (R(T)) of the Bi$_2$Te$_3$ NR device (Fig. 2b) shows a metallic behavior (R decreases with decreasing T), as previously observed in highly doped samples with bulk-dominated conduction[9]. However, for $V_g$ = -6.5V and -15V, we observe an insulating behavior (R increases with decreasing T, due to bulk carrier freeze-out) for 10 K < T < 30K with R saturating at a $V_g$-dependent value ($R^{sat}$) for T < 10 K. This $R^{sat}$ corresponds to $E_F$ inside the bulk bandgap and the low-T metallic conduction of topological surface states (TSS, see band schematic in the inset of Fig. 2b). For $V_g$ = -22V and -30V ($V_g$ < $V_{CNP}$), in addition to the insulating behavior and low-T $R^{sat}$, we observe R(T) peaks at T ~ 10K. This peak resistance is similar to the peak resistance measured at $V_g$ = $V_{CNP}$ from the field effect (Fig. 2a, see also Fig. S1a, note the peak value can vary slightly after thermal cyclings), and is attributed to the strong enhancement of κ of STO[28] at low T, such that $E_F$ is lowered to cross CNP and further into the bulk valence band (VB) as T decreases (see also Fig. S1b, and its inset demonstrating *T*-dependence of $V_{CNP}$). Our data shows that at low T, we can use $V_g$ to tune $E_F$ (schematically represented with dashed lines in the inset of Fig. 2b) all the way from the conduction band (CB) to the TSS then to the VB. The measured R(T) suggests that bulk carriers in our Bi$_2$Te$_3$ NRs can be suppressed with $V_g$ to realize a bulk insulating regime where surface can contribute dominantly to the conduction, and this is further corroborated with magnetotransport measurements presented later. A bulk metallic-to-insulator transition has also been observed in our Bi$_2$Te$_3$ NRs fabricated on SiO$_2$/Si (eg., Fig.

S2, device #4), while we have focused more on devices with STO backgate as they are found to generally give stronger gate tuning for ambipolar field effect.

*Aharonov-Bohm Oscillations.* When an external magnetic field (B) is applied along the length of the NR, the low-T magnetoconductance (G(B)) displays periodic oscillations (Fig. 3a) in B, commonly known as Aharonov-Bohm (A-B) oscillations (or "h/e oscillations") with a characteristic period ($\Delta B = \Phi_o/A$), where $\Phi_o = h/e$ is the flux quantum, A is the NR cross-sectional area (width*height), h is Planck's constant and e is the electron charge. In contrast, for bulk carriers there are impurity-dependent loops with no well-defined cross sectional area, resulting in universal conductance fluctuations (UCF), with non-periodic B-field dependence. A-B oscillations were previously observed in $Bi_2Te_3$[21] and $Bi_2Se_3$[29] NRs. In order to probe the surface states, we measure G vs. parallel B-field from device #1 (Fig. 3a) at four representative $V_g$'s. For $V_g$ = -10, -11 and -13V, periodic oscillations are clearly observed in Fig. 3a with $\Delta B \sim$ 1T (vertical dashed lines). This $\Delta B$ agrees well with $\Phi_o/A$ = 1.06 T for our device, therefore these oscillations are the A-B oscillations, with G maxima at integer multiples of $\Phi_o$ (top axes). However for $V_g$ = 4V ($E_F$ inside CB), the oscillations become non-periodical and they are attributed to UCFs due to bulk carriers.

The FFT of G (with a polynomial background subtracted) vs. 1/B for $V_g$ = -10V and T = 0.3K is depicted in the left inset of Fig. 3a. We observe a dominant peak (~0.98$T^{-1}$), which is the period of the A-B oscillations. This period corresponds to an area A ~ 4000 $nm^2$, again in good agreement with the AFM-measured cross-sectional area of the NR (~3900 $nm^2$) of device#1 (Fig. 1d). The right inset of Fig. 3a displays the T-dependence of the amplitude (from FFT) of the A-B oscillations for $V_g$ = -10V, where a $T^{-1/2}$ dependence is observed (other $V_g$'s show similar behavior). Such a $T^{-1/2}$ dependence is consistent with previous work in TI NRs[21,29] and has also been observed in diffusive metallic[30] and semiconducting rings[31]. We have also observed AB oscillations in two other devices (#2-3) with different cross sectional areas (A = 19800 and 3600 $nm^2$ respectively). The measured period $\Delta B$ vs. $V_g$ are depicted in Fig. 3b, along with the data from device#1. We observe that $\Delta B$ is independent of $V_g$, but instead is controlled by A such that the product $\Delta B \cdot A$ (magnet flux corresponding to $\Delta B$) is ~ h/e (flux quantum) for all 3 devices (inset of Fig. 3b), confirming these oscillations are A-B oscillations of surface carriers.

*Shubnikov-de Haas oscillations.* Figure 4a shows the magneto resistance (R(B)) vs. B-field (perpendicular to the NR surface and the substrate), measured for device #1 at T = 1.5K and $V_g$ = -8V, where small oscillations are observed. The top inset of Fig. 4a displays $\Delta R$ (R(B) with a smooth polynomial background subtracted) vs. 1/B, where periodical oscillations in 1/B are clearly observed. These oscillations are Shubnikov-de Haas (SdH) oscillations due to the formation of Landau levels (LL) in high magnetic fields. We take the FFT of ΔR (bottom inset of Fig. 4a) to find the frequency of oscillations ($B_F$) ~ 55.4T. The Fermi momentum ($k_F$) and carrier density ($n_{SdH}$) can be obtained from $1/B_F = 4\pi e/(hk_F^2)$ and $n_{SdH} = gk_F^2/4\pi$ respectively, where g = 1 is the degeneracy for spin polarized surface state (we attribute the SdH as coming from one surface, most likely the back-gated bottom surface, whereas another surface has much lower mobility and less conduction [9,10,32]). For the data shown in Fig. 4a, we obtain $n_{SdH} = 1.27 \times 10^{12} cm^{-2}$ and $k_F$ = 0.028 Å$^{-1}$. We extract the phase of the SdH oscillations (which will correspond to a Berry's phase) following the standard procedure by plotting the integer Landau level index (n) vs. 1/B (Landau level fan diagram), as shown in Fig. 4b for several $V_g$'s. An integer n is defined at ΔR minima (open markers) and n + 0.5 is defined at ΔR maxima (solid markers). It is clearly seen that n fits (shown as lines) linearly with 1/B ($n = B_F/B + \beta$), the slope ($B_F$) varies with $V_g$ and we extrapolate β ~ 0.5 for all $V_g$'s (right inset of Fig. 4b, noting SdH oscillations were observed in the range of $V_g$ from -8 to -13V for n-type TSS carriers). This β ~ 0.5 is further confirmed by a direct fitting (an example shown as the black solid line in Fig. 4a top inset) of the SdH oscillations to the low-T theoretical expression $\Delta R \sim \exp(-\pi/\mu_{surface}^{SdH} B)\cos(2\pi(B_F/B + 0.5 + \beta))$ (yielding also $B_F$ and SdH mobility $\mu_{surface}^{SdH}$ consistent with values from FFT/LL analysis above and SdH analysis discussed below). We have also plotted $B_F$ ($\propto n_{SdH}$) vs. $V_g$ in the left inset of Fig. 4b, where a linear dependence is observed, demonstrating the effect of gating. The LL indices for both Schrodinger and Dirac fermions can be linearly fitted to 1/B. However, it is well known that β = 0 for Schrodinger fermions and β = 1/2 for massless Dirac fermions, with the Berry's phase given by 2π β [13,33-35]. Such a prominent 1/2-shifted SdH effect provides a strong transport evidence of the spin-helical topological surface state Dirac fermions with the nontrivial Berry's phase π.

The amplitude of the SdH oscillations decreases with increasing T (while at each $V_g$, we found that $n_{SdH}$ does not change substantially with T in the range of temperature where SdH is observed). By fitting the T-dependence of the SdH oscillation amplitudes to the Lifshitz-Kosevich formula (Methods, and an example in Fig. 4c inset) we extract the cyclotron frequency ($\omega_C = eB/m*$) at the B-field position of a given $R_{xx}$ minimum, corresponding to the N$^{th}$ LL. At each $V_g$ (with corresponding $n_{SdH}$), the effective mass (m*) is obtained from the linear fit of $\omega_C$ vs. B-field. The extracted m* vs. $n_{SdH}$ (Fig. 4c) is not constant, rather proportional to $\sqrt{n_{SdH}}$, a distinct character for massless Dirac fermions previously demonstrated in graphene [34,35]. The calculated Fermi velocity ($v_F = \hbar k_F/m*$) is nearly independent of $n_{SdH}$ (inset of Fig. 4d), with a mean value ($\langle v_F \rangle \sim 5 \times 10^5 m/s$), in reasonable agreement with $v_F$ measured by ARPES in Bi$_2$Te$_3$ bulk crystals[6] ($v_F = 4.1 \times 10^5 \, m/s$). Figure 4d shows the corresponding Fermi energy ($E_F = m^* \langle v_F \rangle^2$) as well as the m* vs. $k_F$, where a linear dependence of $E_F$ (and m*) with $k_F$ is demonstrated (dashed line represents a calculated linear dependence $E_F = m^* v_F^2 = \hbar v_F k_F$ using $v_F = \langle v_F \rangle$ as constant Fermi velocity).

The density-independent $v_F$, gate-tunable m* ~ $\sqrt{n_{SdH}}$ and therefore the linear band dispersion ($E_F$ vs. $k_F$) obtained in our experiments are in good agreement with the linear energy dispersion measured by ARPES in bulk Bi$_2$Te$_3$[6,7], and are clear evidences that carriers in Bi$_2$Te$_3$ NRs behave as massless Dirac fermions in electronic transport.

We have extracted the (high-magnetic-field) quantum lifetime ($\tau_q$) of surface carriers for each $V_g$ (Fig. S3 shows 2 examples) from the slope of $\ln(\Delta RB \sinh(2\pi^2 k_B T / \hbar \omega_C))$ vs. 1/B (so called "Dingle analysis", see Methods and Supplemental Materials). We also calculated the corresponding high-field quantum mean free path ($\ell_q = v_F \tau_q$), SdH mobility $\mu_{surface}^{SdH} = \dfrac{e\tau_q}{m^*}$, and SdH conductivity $\sigma_{surface}^{SdH} = e n_{Sdh} \mu_{surface}^{SdH}$ (See Fig. S4) of the surface carriers that give SdH oscillations at T = 1.5K. Since it is expected that $\tau_q < \tau_{tr}$, where $\tau_{tr}$ is the transport lifetime (contributed mainly by large-angle scattering rather than by all scatterings as in $\tau_q$) [36] of surface state carriers at B = 0T, the calculated $\mu_{surface}^{SdH}$ and $\sigma_{surface}^{SdH}$ provide *lower bounds* for the (zero-field) transport mobility ($\mu_{surface}$) and conductivity ($\sigma_{surface}$) of the corresponding surface. We have plotted the (dimensionless) ratio $\sigma_{surface}^{SdH} / \sigma_{total}$ vs. the gate voltage in Fig. S5, where $\sigma_{total} = (L/W)/R$ is the total (2D sheet) conductivity of the NR (measured at zero-magnetic-field, with L, W and R being the length, width and 4-terminals resistance of the TI NR, respectively; noting all conductivities in this paper are 2D sheet conductivities). For the reasons discussed above (due to $\tau_q < \tau_{tr}$) and the fact that the contribution from only one surface (that gives SdH oscillations) is included in $\sigma_{surface}^{SdH}$, this ratio ($\sigma_{surface}^{SdH} / \sigma_{total}$) *underestimates* the actual surface contribution to the total conductance. The large value of $\sigma_{surface}^{SdH} / \sigma_{total}$ (reaching ~0.5 at $V_g$ ~ -8V) is notable, confirming that the surface can make substantial or even dominant contribution (manifested in the residual saturating metallic resistance seen in the R vs T in Fig. 2b) to the total conductance at low T (where bulk is insulating at this $V_g$). The reduction of $\sigma_{surface}^{SdH} / \sigma_{total}$ for decreasing $V_g$ (towards $V_{CNP}$) might be related to the shape of the bandstructure of $Bi_2Te_3$ near the Dirac point (DP), which is buried inside the bulk VB (Fig. 2b inset).

*Weak antilocalization and localization.* Weak antilocalization (WAL) appears as a negative magnetoconductivity ($\Delta\sigma(B) = \sigma(B) - \sigma(B = 0T)$) with a sharp cusp centered at zero B-field. It is commonly observed in TIs due to the spin-momentum locking resulting from strong spin-orbit coupling. The π-Berry's phase carried by the topological surface states of TIs leads to destructive interference between time-reversed paths, enhancing $\sigma(B=0T)$. The WAL cusp can be destroyed by applying a B-field that breaks the time reversal symmetry and π-Berry's phase[37]. We use the simplified Hikami-Larkin-Nagaoka formula[37,38] to fit $\Delta\sigma(B)$:

$$\Delta\sigma_{HLN}(B) = \alpha_0 \frac{e^2}{\pi h}\left[\Psi\left(\frac{\hbar}{4eBL_{\phi 0}^2}+\frac{1}{2}\right) - \ln\left(\frac{\hbar}{4eBL_{\phi 0}^2}\right)\right] \quad (1)$$

where $\Psi$ is the digamma function, $L_{\phi 0}$ is the phase coherence length, and $\alpha_0$ is a prefactor. Each coherent channel that carries a π-Berry's phase should give $\alpha_0 = -1/2$. In contrast, if the Berry's phase is absent, the destructive interference mentioned above would become constructive, resulting in weak localization (WL). The WL appears as a positive magnetoconductivity that can be similarly fitted to Eq. (1) with a positive $\alpha_0$. Figure 5a shows the magnetoresistance R(B) measured at 1.5 K at five different $V_g$'s. By increasing $V_g$ from -18V to +14V, we observe that R(B) transitions from exhibiting a dip (indicating negative $\Delta\sigma(B)$, and WAL) to a peak (indicating positive $\Delta\sigma(B)$, and WL) near zero B field. Figure 5b displays the anti-symmetrized $\Delta\sigma(B)$ (extracted from the R(B)) vs. B-field at these $V_g$'s, where the gate induced transition from negative to positive $\Delta\sigma(B)$ (WAL to WL) is clearly observed. Also, we note that the amplitude of the WAL or WL cusp depends on $V_g$. We first focus on the gate-tunable WAL for $V_g < 0V$. Using equation 1, we fit $\Delta\sigma(B)$ (for B < 2T) at different $V_g$'s and extract the gate-dependent $\alpha_0$ and $L_{\phi 0}$ (Fig. 5c). We observe $\alpha_0$ peaks ~ -1 for $V_g$ ~ -18 V (~$V_{CNP}$) and approaches -0.5 for either $V_g > -12V$ ($E_F$ approaching the CB) or $V_g$ ~ -30V ($E_F$ near the VB). The parameter $\alpha_0$ is related to the number of coherent conduction channels $A = 2|\alpha_0|$. The transition from $\alpha = -0.5$ (A = 1, away from CNP) to $\alpha = -1$ (A = 2, near CNP) tuned by $V_g$ reflects the transition of the system from a single coherent channel (with bulk and TSS strongly coupled together) to two decoupled coherent channels surrounding the two surfaces (as bulk conduction is suppressed by the gate [23,38]). Also, $L_{\phi 0}$ displays a minimum of ~ 62 nm for $V_g$ ~ $V_{CNP}$. For $V_g \geq 2V$, $\Delta\sigma(B)$ transitions into positive and weak localization (WL) behavior, although a small WAL (negative $\Delta\sigma(B)$) can still be observable at very low B fields (Fig. 5b), reflecting a competition (mixed contributions) between WAL and WL. We use the two-component HLN formula [39-41] to fit Δσ(B) and extract the WAL ($\alpha_0$, $L_{\phi 0}$) and WL ($\alpha_1$, $L_{\phi 1}$) contributions,

$$\Delta\sigma_{HLN}(B) = \sum_{i=0,1} \frac{\alpha_i e^2}{\pi h}\left[\Psi\left(\frac{\hbar}{4eBL_{\phi i}^2} + \frac{1}{2}\right) - \ln\left(\frac{\hbar}{4eBL_{\phi i}^2}\right)\right] \quad (2).$$

In order to obtain unique fitting results for the WAL and WL parameters (using equation 2), we fix the WAL prefactor $\alpha_0$ = -0.5 (a reasonable choice given Fig. 5c shows $\alpha_0$ saturates at ~-0.5 for $V_g$ >-12V). We observe (Fig. 5d) that as $V_g$ increases from 2V to 14V, both the WL prefactor $\alpha_1$ (~1) and WAL phase coherence length $L_{\Phi 0}$ (~ 120 nm) remain relatively constant, while the WL phase coherence length $L_{\Phi 1}$ moderately increases from 46 to 76 nm. Transitions from WAL to WL in TIs have previously been observed in magnetically doped TIs[40] and extremely thin TI flakes (4-5 nm thick) [39,42], and are attributed in both cases to gap opening in the surface states (i.e., DP destroyed) [41,43]. It has been theoretically predicted that TI can also display a competition between WAL and WL, *without* involving surface gap opening and destroying the DP, if $E_F$ is close to the transition region between TSS and the bulk bands (CB or VB)[22,23]. It has been pointed out that as $E_F$ moves from the TSS to near the bottom of CB, the Berry's phase decreases from of π to 0 (due to change in the pseudospin texture, where the bulk bandgap can act as a fictitious Zeeman field perpendicular to the spin-orbit field), giving rise to a WAL to WL transition [22,23]. Observing such WAL to WL transition as predicted[22,23] is another piece of evidence that we are actually observing topological surface states transport.

*Conclusions*

In conclusion, we have achieved ambipolar field effect on single crystal $Bi_2Te_3$ NRs using STO as substrate and backgate dielectric. We are able to use the gate to tune the electronic transport from being dominated by the metallic bulk to a bulk-insulating regime such that surface state contributes mostly to the conduction at low temperatures. We observed gate-controlled A-B oscillations of surface carriers. We have also measured gate-controlled SdH oscillations with an extracted π-Berry's phase, m* ~ $\sqrt{n_{2D}}$, and nearly constant $v_F$, providing direct evidence of the Dirac fermion nature (linear $E_F$ vs. $k_F$) of the topological surface states. Also, a gate-tunable WAL is observed and the extracted number of coherent conduction channels is found to peak around 2 (corresponding to 2 decoupled channels surrounding 2 surfaces) for $V_g$ ~ $V_{CNP}$. Finally, we observe a competition between WAL and WL for $V_g$ > 0V, consistent with a predicted collapse of Berry's phase and change of pseudospin texture as $E_F$ approaches the CB. The gate-tunable topological surface states found in our samples pave the way towards topologically protected nanoelectronic and spintronic devices and possible applications in topological quantum computing with Majorana fermions.

**Methods**

*Nanoribbons synthesis and transfer.* Bulk $Bi_2Te_3$ (99.999%, Alfa Aesar) was placed in the center of a single-zone furnace (TF55035A-1, Lindburg® BlueM®) and $SiO_2$ covered Si substrates were placed downstream. The temperature was slowly ramped (1–2 $^oC$/min) to ~480 $^oC$ under flowing Ar (~75 $cm^3$/min, 99.999% purity) and pressure was maintained in the range of 20–70 Torr for the duration of the synthesis. Growth times were on the order of 1–3 hours.

*Transmission electron microscopy (TEM).* Transmission electron microscopy (200 kV, JEM-2010F, JEOL Ltd.) analysis of several NRs showed growth along the $[11\bar{2}0]$ direction and energy dispersive X-ray spectroscopy indicated excess tellurium (66±2 at.% Te). The tips of the NRs were observed to be catalyst free with $\{10\bar{1}0\}$ facets.

*Raman Spectroscopy.* Raman spectroscopy was performed using a Horiba Jobin Yvon Xplora confocal Raman microscope. The wavelength of the excitation laser was 532 nm and the power of the laser incident on the sample was kept below 200 µW to avoid sample burning. The laser spot size was ~ 0.6 µm with a 100X objective lens (numerical aperture = 0.90). The spectral resolution was 1.0 $cm^{-1}$ (using a grating with 2400 grooves/mm) and each spectrum was an average of 3 acquisitions (5 seconds of accumulation time per acquisition).

*Atomic Force Microscopy (AFM).* The width and thickness of the samples was measured by a NT-MDT NTEGRA Prima multifunctional atomic force microscope in tapping mode configuration. In order to avoid any damage or NRs displacement, we use a small cantilever driving amplitude and set point amplitude corresponding to 75% of the freely resonant magnitude. The silicon probes used are ACTA series from AppNANO, with a nominal 40 N/m spring constant, cantilever length of 125 µm and 6 nm radius of curvature.

*Device fabrication and electronic transport measurements.* Our STO substrates are single side polished (100) from Shinkosha Ltd. The electrical contacts (Cr/Au, 5nm/70nm, e-beam evaporated) to $Bi_2Te_3$ NRs were patterned by e-beam lithography. Right before the metal deposition, a short etch (20 sec.) in a dilute solution of sulfuric acid ($H_2SO_4$:$H_2O$, 1:10) is performed to remove native oxide from the surface. After the metal evaporation, the device is mounted and wirebonded to a ceramic chip carrier. Cr/Au (10/70 nm) is deposited on the backside of STO (backgate electrode). All the electrical measurements are performed in a top loading Helium-3 system (Heliox TL system, Oxford Instruments). The resistances were measured by 4-terminals method and using low frequency lock-in detection (PAR124A), with a driving current of ~1μA.

*Temperature dependence of the Shubnikov de Haas oscillations and effective mass calculation.* We fit the T-dependence of SdH oscillation amplitude (ΔR) to the Lifshitz-Kosevich theory[44]:

$$\Delta R(T,B) \propto \frac{2\pi^2 k_B T / \hbar \omega_C}{\sinh[2\pi^2 k_B T / \hbar \omega_C]} e^{-\pi/\omega_C \tau_q} \quad (3)$$

Where $\omega_C$ and $\tau_q$ are the fitting parameters and B is the magnetic field position of the $N^{th}$ minimum in ΔR; $k_B$ is the Boltzmann's constant; $\omega_C = eB/m^*$ is the cyclotron frequency (and $\hbar \omega_C$ is the cyclotron energy gap for the $N^{th}$ LL); m* is the effective mass of the carriers and $\tau_q$ is the quantum lifetime. The $\omega_C$ is calculated from the fitting of the relative amplitude ΔR/R(B = 0T) as a function of T using equation (3) (inset of Fig. 4c) for different LLs. The m* is extracted from the slope of $\omega_C$ vs. B for a given $V_g$.

**Acknowledgements**

The TI material synthesis, characterization and magneto-transport studies are supported by DARPA MESO program (Grant N66001-11-1-4107). The FET fabrication and characterizations are supported by Intel Corporation. L. A. J. acknowledges support by an Intel PhD fellowship. L.P.R. acknowledges support by DOE grant DE-SC0008630.

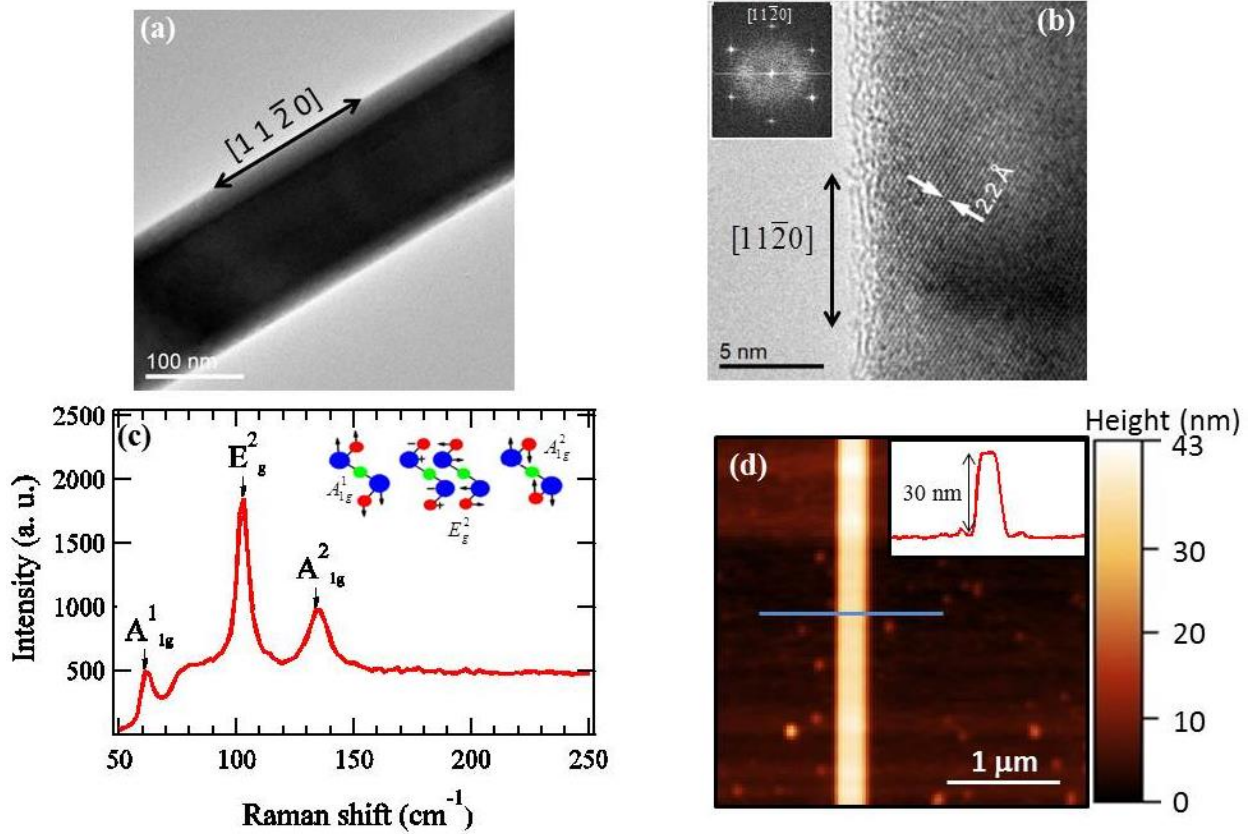

Figure 1. **Material characterizations of $Bi_2Te_3$ nanoribbons.** (a) Transmission electron microscope (TEM) image of a 150 nm wide $Bi_2Te_3$ nanoribbon (NR). NRs grow along the $[11\bar{2}0]$ direction. (b) High-resolution TEM image, with the corresponding Fourier transform depicted in the inset. The obtained lattice spacing of 2.2 Å is consistent with the lattice spacing of $\{11\bar{2}0\}$ planes [16,24]. (c) A representative Raman spectrum (measured with a 532nm laser) showing characteristic Raman peaks (labeled) similar to those observed in bulk $Bi_2Te_3$ (inset depicts corresponding phonon modes [25,26]) (d) Atomic force microscope (AFM) image of a 130nm wide $Bi_2Te_3$ NR (device #1 before fabrication) on a $SrTiO_3$ (STO) substrate. A thickness of 30nm is extracted from the AFM line profile (inset, measured along the blue line in the AFM image).

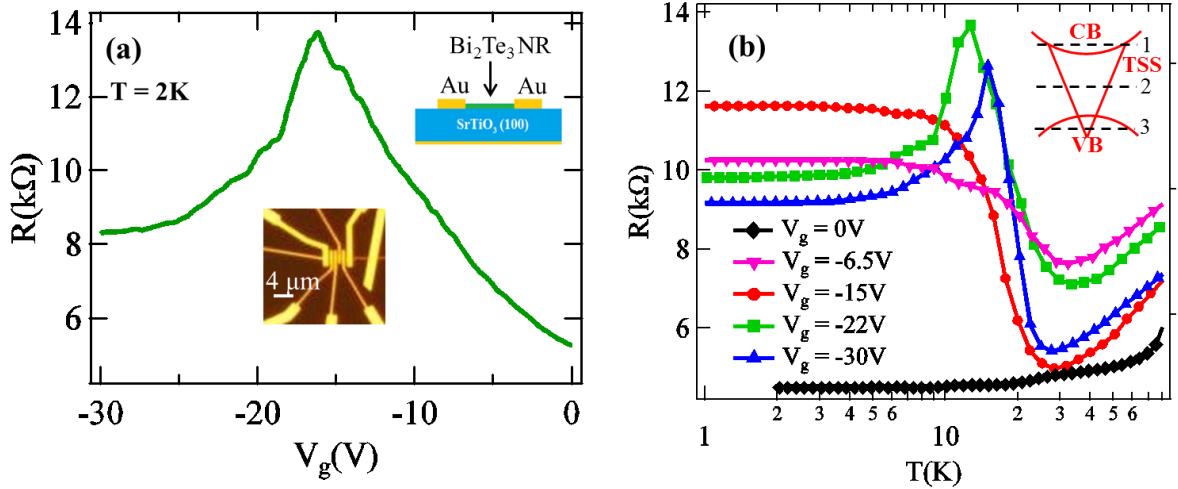

Figure 2. **Gate-tuned bulk metal-insulator transition and ambipolar field effect.** (a) Ambipolar field effect (4-terminals resistance R vs. $V_g$), measured at T = 2K. The upper (lower) inset depicts a schematic cross section (optical image, top view, with the inner electrodes used to measure the voltage separated by 850 nm) of the fabricated $Bi_2Te_3$ NR device #1 on a 500 μm-thick STO substrate. (b) Temperature (plotted in log scale) dependence of R of device #1 measured at five gate voltages ($V_g$). Inset: schematic of the band diagram of $Bi_2Te_3$. The horizontal dashed lines depict 3 representative locations of the Fermi energy ($E_F$), intercepting with the bulk conduction band (CB), topological surface states (TSS) and bulk valence band (VB), respectively. Note the Dirac point (DP) is buried inside the VB, thus only n-type TSS carriers are accessible inside the bulk bandgap.

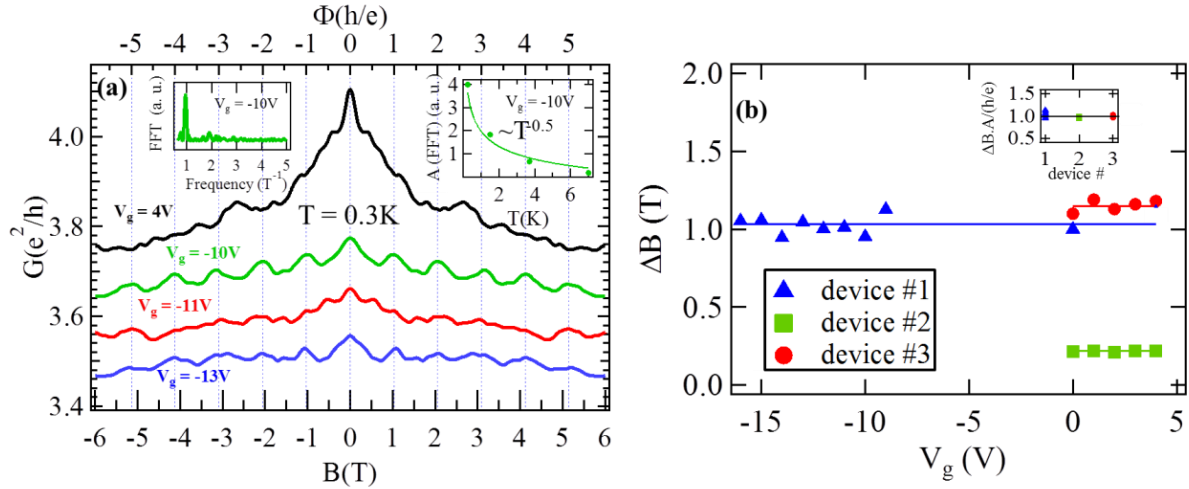

Figure 3. **Aharonov-Bohm oscillations of the surface carriers.** (a) Magnetoconductance (G) vs. magnetic field (B, with the corresponding magnetic flux Φ plotted on the top axis) applied parallel to the NR length (device #1) at 4 different $V_g$'s (curves offset for clarity, except for $V_g$ = -13V) measured at T = 0.3 K. Left inset: FFT of G ($V_g$ = -10V) vs. 1/B, after a background subtraction in G. Right inset: temperature dependence of the FFT amplitude for the h/e oscillations at $V_g$ = -10V; a fitting with a $T^{-0.5}$ dependence is plotted as a solid line. (b) Period (ΔB) of oscillations vs. $V_g$ for three devices with different cross section areas (A = 3900, 19800 and 3600 nm$^2$ for devices 1, 2 and 3 respectively). The inset depicts magnetic flux (ΔB•A, in units of h/e) for these 3 devices (including all data points measured at different $V_g$'s shown in the main panel).

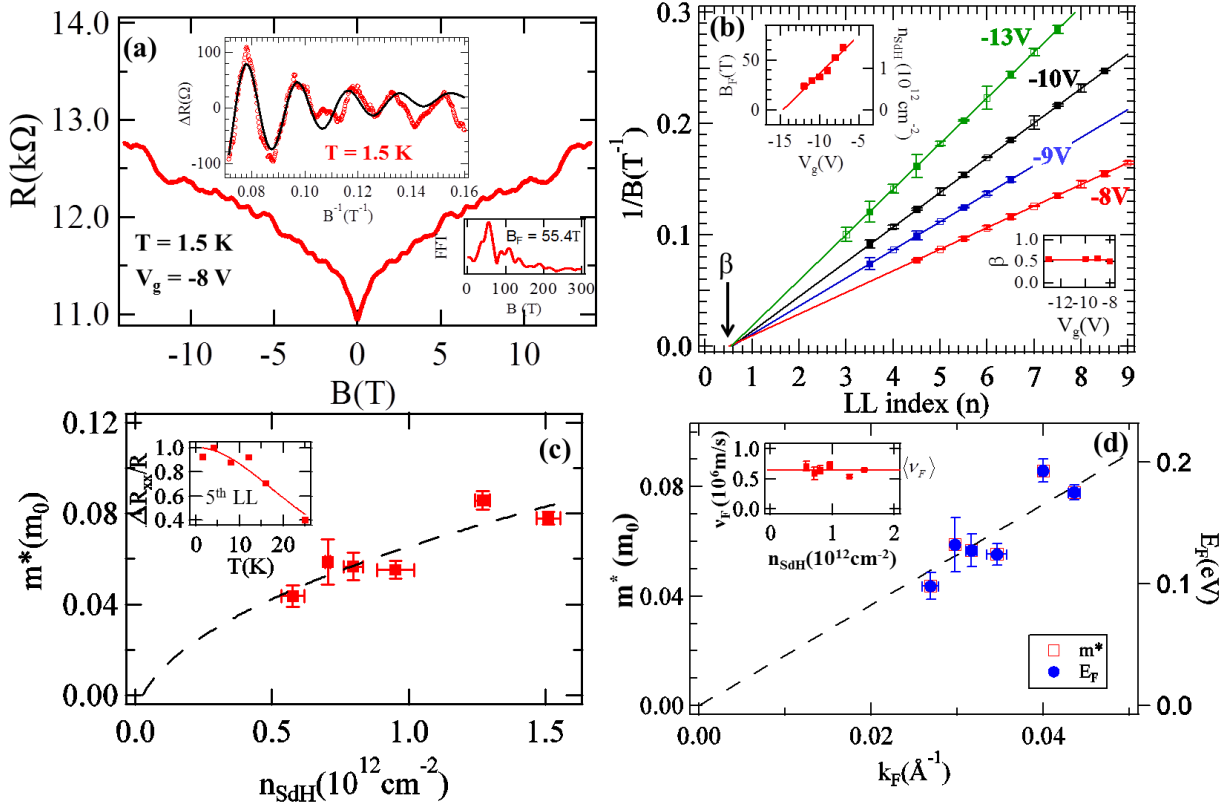

Figure 4. **Shubnikov-de Haas (SdH) oscillations of surface state Dirac fermions showing linear E-k dispersion.** (a) Representative magnetoresistance (R vs. B) with B applied perpendicular to the NR plane, measured at $V_g$ = -8V and T = 1.5 K for device #1. Top inset: ΔR (R with a background substracted) vs. 1/B (red, with a fit shown as black curve and FFT shown as bottom inset). (b) 1/B vs. LL index (fan diagram) at different $V_g$'s. Right inset: β (intercept of LL index for 1/B →0 as extracted from the fan diagram) vs. $V_g$. Left inset: SdH oscillation frequency ($B_F$, inverse slope of 1/B vs. LL index) vs. $V_g$ (solid line is a linear fit) (c) Extracted effective mass (m*) vs. carrier density ($n_{SdH}$ = $eB_F/h$), where m* is extracted by equation (3) from the temperature dependence of the SdH oscillations. The dashed line is a fit to $\sqrt{n_{SdH}}$ behavior. The inset shows ΔR/R vs. T, for LL index = 5. (d) Effective mass m* (left axis) as well as Fermi energy ($E_F = m^* \langle v_F \rangle^2$) vs. Fermi momentum ($k_F = \sqrt{4\pi eB_F/h}$), where $\langle v_F \rangle$ is the average (and mostly constant) Fermi velocity ($v_F = \hbar k_F/m^*$) over the $n_{SdH}$ range measured (inset). Black dashed line represents a calculated linear dependence using $\langle v_F \rangle$ as Fermi velocity.

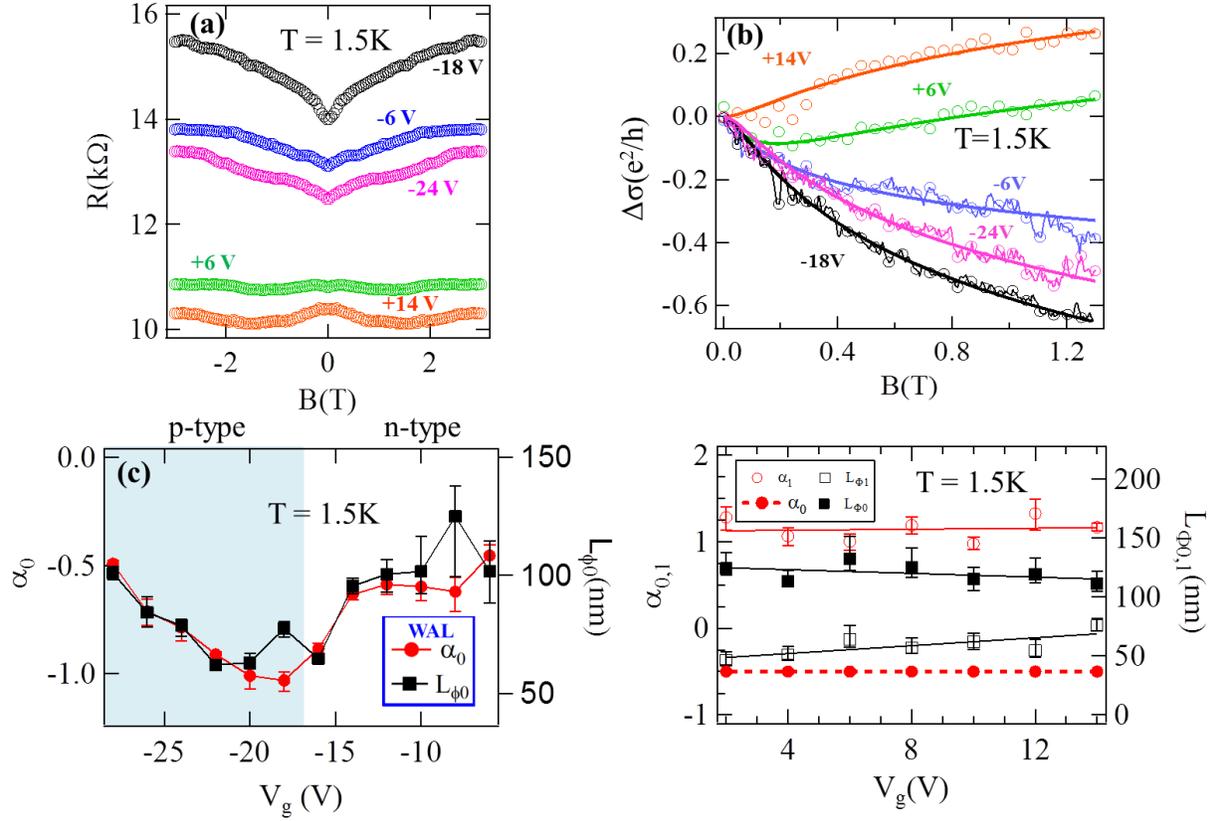

Figure 5. **Gate-tuned weak antilocalization (WAL) and transition to weak localization (WL).** **(a)** Magnetoresistance (R vs. B, with B applied perpendicular to the NR plane) at 5 different $V_g$'s at T = 1.5 K for device #1. **(b)** Magnetoconductivity ($\Delta\sigma(B)$, where $\Delta\sigma(B) = \sigma(B) - \sigma(B = 0T)$), normalized by $e^2/h$. Solid lines depict fitting to HLN equation (see text). **(c)** Weak antilocalization (WAL) prefactor $\alpha_0$ (left axis) and phase coherence length $L_{\Phi 0}$ (right axis), both extracted using the HLN fit, as functions of $V_g$ (for negative $V_g$). **(d)** Weak localization (WL) prefactor $\alpha_1$ (left axis) and phase coherence lengths (right axis) for WAL and WL contributions (solid lines are guides to the eye), extracted from the two-component HLN fit, with fixed $\alpha_0 = -0.5$, for $V_g > 2V$.

**Supplementary Information**

Figure S1a depicts the field effect of R at T=0.3K for device #1 in different cool downs. The measurements were in the following order: black, blue and green, measured after measuring R vs. T for $V_g$ = 0, -30 and -22V respectively. The magenta dashed curve was a repeat measurement for the green curve to show the reproducibility of R vs. $V_g$ in the same cool down. We observe that $V_{CNP}$ and the peak resistance have moderate changes with different cool-downs (which may reflect re-arrangement of impurities inside TI and the STO substrate). Figure S1b depicts the field effect of R for device #1 (main text) at different T's. The $V_{CNP}$ changes from $V_{CNP}$ ~ -15V (T = 4K) to $V_{CNP}$ ~ -40V (T = 30K). The change of $V_{CNP}$ with temperature can be attributed to the significant reduction of STO gate capacitance as temperature increases from 4K to 30K [28].

Figure S2, shows the R(T) for another TI NR device (device #4) on $SiO_2$ (300nm thick)/doped Si substrate (with gate capacitance largely T-independent) for 2 different $V_g$'s. Data at $V_g$ = 0V show a metallic behavior (R decreases with decreasing T). However, for $V_g$ = -70V, we observe an insulating behavior (R increases with decreasing T) for T > 70 K with R saturating ($R^{sat}$) for T < 70 K. This corroborates the gate-tuned metal to insulator transitions in the bulk of $Bi_2Te_3$ NRs in devices fabricated on STO substrate (Fig. 2b).

*Extraction of quantum lifetime, SdH mobility and SdH conductivity*

Figure S3a (S3b) displays $\ln(\Delta RB\sinh(2\pi^2 k_B T/\hbar\omega_C))$ vs. 1/B at $V_g$ = -8V ($V_g$ = -12V) at T = 1.5K. The $\tau_q$ is extracted from the slope of $\ln(\Delta RB\sinh(2\pi^2 k_B T/\hbar\omega_C))$ vs. 1/B, where $\tau_q$ varies from 0.5 x10$^{-13}$ sec ($V_g$ = -12V) to 1.3x10$^{-13}$ sec ($V_g$ = -8V) in our experiments. The surface mobility ($\mu_{surface}^{SdH} = \frac{e\tau_q}{m*}$) is found to be $\mu_{surface}^{SdH}$ ~ 2,000 – 3,000 cm$^2$/Vs and quantum mean free path ($\ell_q = v_F \tau_q$) ~ 30 - 50 nm, as depicted in Fig. S4. The gate-dependent ratio between high-B surface SdH conductivity ($\sigma_{surface}^{SdH} = en_{Sdh}\mu_{surface}^{SdH}$, which is smaller than the surface transport conductivity at zero-B) and total conductivity ($\sigma_{total}$, measured at zero-B) is plotted in Fig. S5, showing a significant value with notable enhancement near CNP, consistent with the expected large surface to bulk conductance ratio (which is bounded from below by the plotted ratio) near CNP.

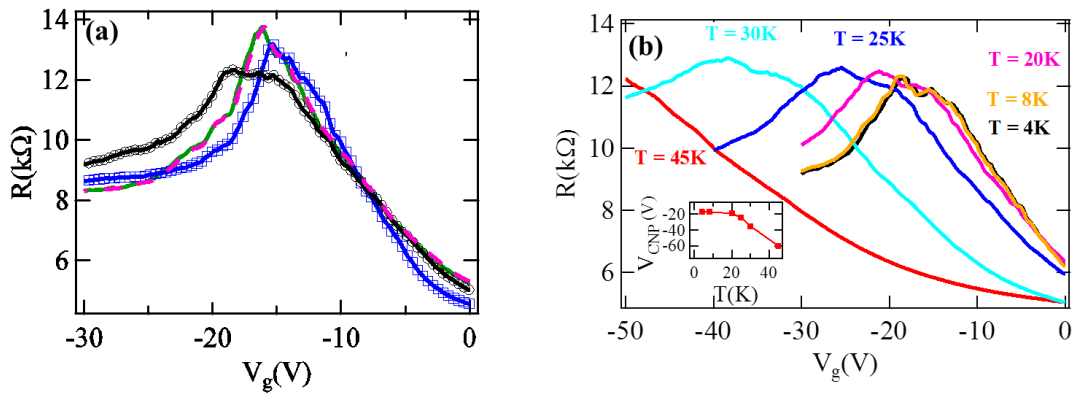

Figure S1. **Field effect and T-dependence of $Bi_2Te_3$ NR on STO (device #1).** (a) R vs. $V_g$ measured for device #1 at T = 0.3 K, each curve was taken after the thermal cycling caused by the measurements of R vs. T depicted in the main Fig. 2b. The measurements were in the following order: black, blue and green (magenta), measured after measuring R vs. T for $V_g$ = 0, -30 and -22V respectively. The magenta dashed curve was a repeat measurement of the green curve to show the reproducibility of R vs. $V_g$ in the same cool down. (b) Field effect of Resistance (R) measured at different temperatures for device #1. Inset: temperature dependence of the $V_{CNP}$.

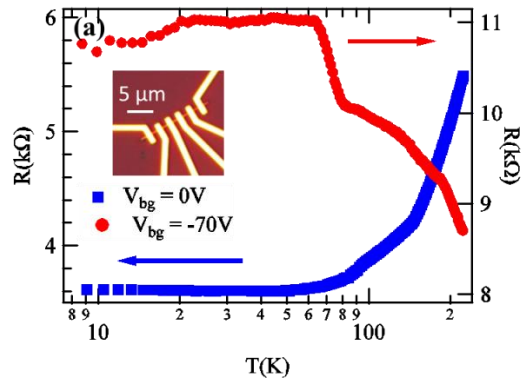

Figure S2. **Temperature dependence of the resistance of a $Bi_2Te_3$ NR on $SiO_2$/Si.** (a) R vs. T measured from device #4 ($Bi_2Te_3$ NR of width = 330nm and thickness ~ 30nm on 300nm $SiO_2$/doped Si) at two $V_g$'s. The inset depicts the optical image of the device.

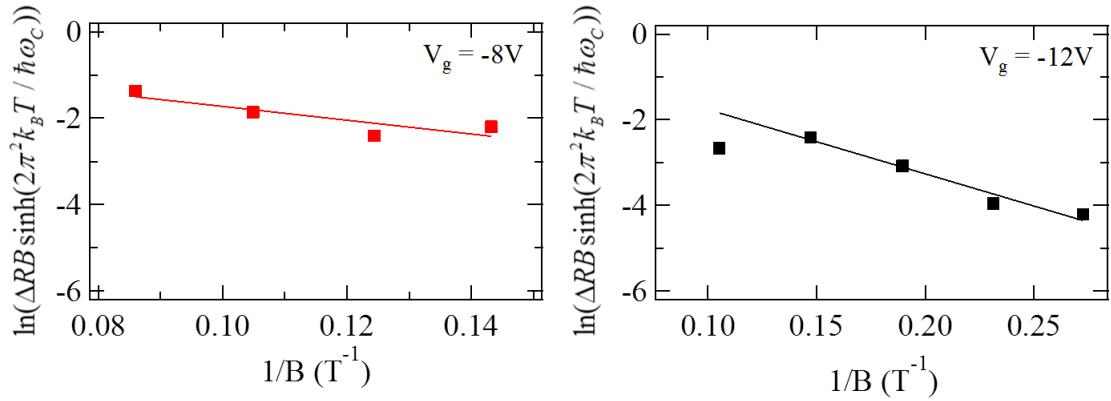

Figure S3. **Representative examples of the Dingle analysis used to obtain the quantum lifetime ($\tau_q$).** $\ln(\Delta RB \sinh(2\pi^2 k_B T / \hbar \omega_C))$ vs $1/B$ at $V_g = -8V$ **(a)** and $V_g = -12V$ **(b)** measured at $T = 1.5K$ for device #1, where $\tau_q$ is obtained from the slope. The $\omega_C$ and $\Delta R$ are defined in Methods.

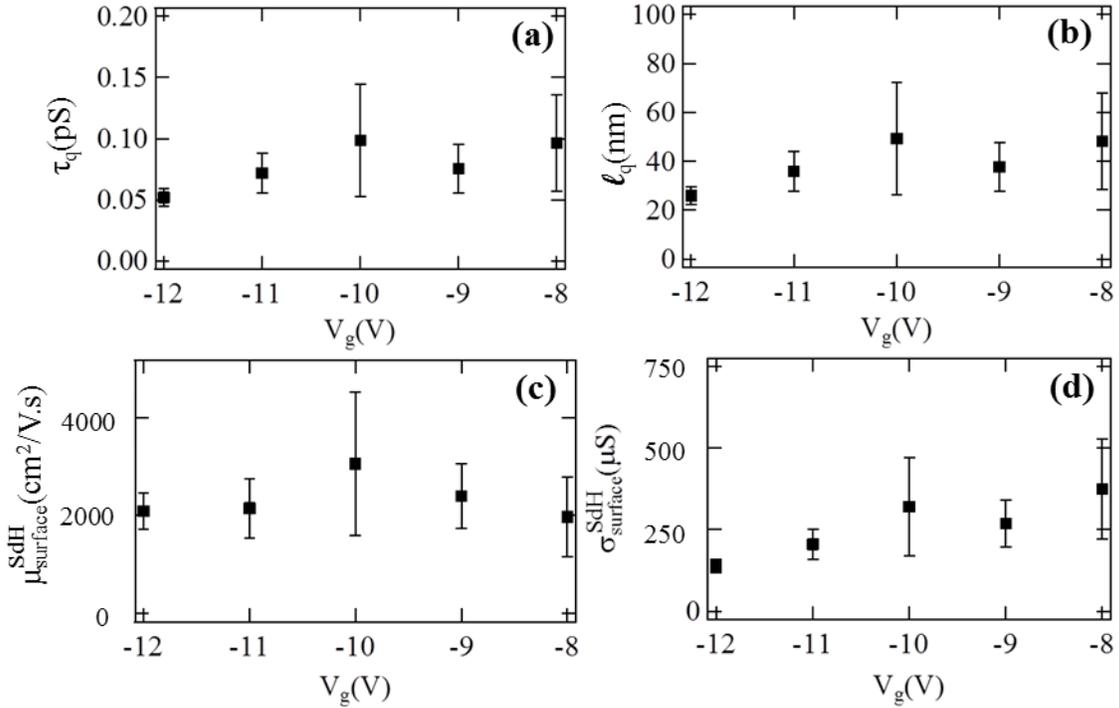

Figure S4. **Surface state quantum lifetime (a), quantum mean free path (b), high-field mobility (c) and high-field conductivity (d) extracted from the SdH oscillations,** extracted for 5 $V_g$'s (for device #1

at T = 1.5K. Note the high-B (SdH) surface mobility ($\mu_{surface}^{SdH}$) and conductivity ($\sigma_{surface}^{SdH}$) are *lower bounds* for the surface mobility and conductivity measured at B = 0T respectively.

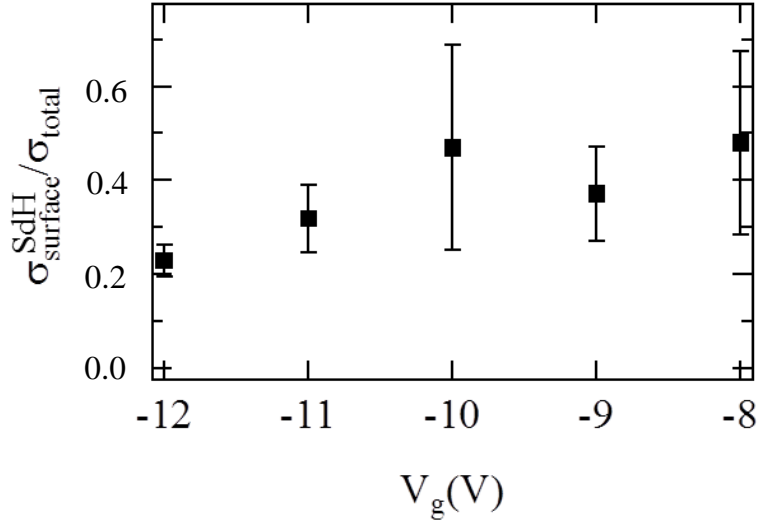

Figure S5. **Ratio of high-field surface conductance to zero-field total conductance.** Shown is $\sigma_{surface}^{SdH}/\sigma_{total}$ at T = 1.5K vs. $V_g$ for device #1, where $\sigma_{surface}^{SdH} = en_{Sdh}\mu_{surface}^{SdH}$, $\ell_q$ is the quantum mean free path ($\ell_q = v_F\tau_q$), $\tau_q$ is the quantum lifetime of the surface states (Fig. S4), and $\sigma_{total}$ is the total conductivity of the NR at B = 0T. The plotted ratio is a lower bound for the surface-to-total conductance ratio at zero-B, as discussed in the main text.

# References

1. Moore, J. E. The birth of topological insulators. *Nature* **464**, 194 (2010).
2. Hasan, M. Z. & Kane, C. L. Colloquium: Topological insulators. *Reviews of Modern Physics* **82**, 3045 (2010).
3. Qi, X.-L. & Zhang, S.-C. Topological insulators and superconductors. *Reviews of Modern Physics* **83**, 1057 (2011).
4. Cook, A. & Franz, M. Majorana fermions in a topological-insulator nanowire proximity-coupled to an s-wave superconductor. *Phys. Rev. B* **84**, 201105 (2011).
5. Fu, L. & Kane, C. Superconducting proximity effect and majorana fermions at the surface of a topological insulator. *Phys. Rev. Lett.* **100**, 96407 (2008).
6. Chen, Y. L. *et al.* Experimental Realization of a Three-Dimensional Topological Insulator, $Bi_2Te_3$. *Science* **325**, 178 (2009).
7. Hsieh, D. *et al.* A tunable topological insulator in the spin helical Dirac transport regime. *Nature* **460**, 1101 (2009).
8. Zhang, T. *et al.* Experimental Demonstration of Topological Surface States Protected by Time-Reversal Symmetry. *Phys. Rev. Lett.* **103**, 266803 (2009).
9. Qu, D.-X., Hor, Y., Xiong, J., Cava, R. & Ong, N. Quantum oscillations and hall anomaly of surface states in the topological insulator $Bi_2Te_3$. *Science* **329**, 821 (2010).
10. Jun, X. *et al.* Quantum oscillations in a topological insulator $Bi_2Te_2Se$ with large bulk resistivity (6Ωcm). *Physica E: Low-dimensional Systems and Nanostructures* **44**, 917 (2012).
11. Kong, D. *et al.* Ambipolar field effect in the ternary topological insulator $(Bi_xSb_{(1-x)})_2Te_3$ by composition tuning. *Nat. Nanotechnol.* **6**, 705 (2011).
12. Ren, Z., Taskin, A., Sasaki, S., Segawa, K. & Ando, Y. Large bulk resistivity and surface quantum oscillations in the topological insulator $Bi_2Te_2Se$. *Phys. Rev. B* **82**, 241306 (2010).
13. Taskin, A., Ren, Z., Sasaki, S., Segawa, K. & Ando, Y. Observation of dirac holes and electrons in a topological insulator. *Phys. Rev. Lett.* **107**, 016801 (2011).
14. Analytis, J. G. *et al.* Two-dimensional surface state in the quantum limit of a topological insulator. *Nature Physics* **6**, 960 (2010).
15. Wang, Z. *et al.* Tuning carrier type and density in $Bi_2Se_3$ by Ca-doping. *Appl. Phys. Lett.* **97**, 042112 (2010).
16. Kong, D. S. *et al.* Topological Insulator Nanowires and Nanoribbons. *Nano Lett.* **10**, 329 (2010).
17. Tian, M. L. *et al.* Dual evidence of surface Dirac states in thin cylindrical topological insulator $Bi_2Te_3$ nanowires. *Sci Rep* **3**, 7 (2013).
18. Hong, S., Cha, J., Kong, D. & Cui, Y. Ultra-low carrier concentration and surface-dominant transport in antimony-doped $Bi_2Se_3$ topological insulator nanoribbons. *Nat. Commun.* **3**, 757 (2012).
19. Steinberg, H., Gardner, D., Lee, Y. & Jarillo-Herrero, P. Surface State Transport and Ambipolar Electric Field Effect in $Bi_2Se_3$ Nanodevices. *Nano Lett.* **10**, 5032 (2010).
20. Yuan, H. *et al.* Liquid-gated ambipolar transport in ultrathin films of a topological insulator $Bi_2Te_3$. *Nano Lett.* **11**, 2601 (2011).
21. Xiu, F. X. *et al.* Manipulating surface states in topological insulator nanoribbons. *Nat. Nanotechnol.* **6**, 216 (2011).
22. Lu, H. Z. & Shen, S. Q. Weak localization of bulk channels in topological insulator thin films. *Phys. Rev. B* **84**, 125138 (2011).
23. Garate, I. & Glazman, L. Weak localization and antilocalization in topological insulator thin films with coherent bulk-surface coupling. *Phys. Rev. B* **86**, 035422 (2012).
24. Kong, D. *et al.* Few-layer nanoplates of $Bi_2Se_3$ and $Bi_2Te_3$ with highly tunable chemical potential. *Nano Lett.* **10**, 2245 (2010).
25. Richter, W., Kohler, H. & Becker, C. R. Raman and far-infrared investigation of phonons in rhombohedral $V_2$-$VI_3$ compounds - $Bi_2Te_3$, $Bi_2Se_3$, $Sb_2Te_3$ and $Bi_2(Te_{1-X}Se_X)_3$ (0 < x <1), $(Bi_{1-Y}Sb_Y)_2Te_3$ (0 < y < 1). *Physica Status Solidi B-Basic Research* **84**, 619 (1977).
26. Shahil, K., Hossain, M., Teweldebrhan, D. & Balandin, A. Crystal symmetry breaking in few-quintuple $Bi_2Te_3$ films: Applications in nanometrology of topological insulators. *Appl. Phys. Lett.* **96**, 153103 (2010).
27. Teweldebrhan, D., Goyal, V. & Balandin, A. A. Exfoliation and Characterization of Bismuth Telluride Atomic Quintuples and Quasi-Two-Dimensional Crystals. *Nano Lett.* **10**, 1209 (2010).

28  Couto, N., Sacépé, B. & Morpurgo, A. Transport through graphene on SrTiO3. *Phys. Rev. Lett.* **107**, 225501 (2011).
29  Peng, H. L. *et al.* Aharonov-Bohm interference in topological insulator nanoribbons. *Nat. Mater.* **9**, 225 (2010).
30  Washburn, S., Umbach, C. P., Laibowitz, R. B. & Webb, R. A. Temperature-Dependence of the Normal-Metal Aharonov-Bohm Effect. *Phys. Rev. B* **32**, 4789 (1985).
31  Hansen, A. E., Kristensen, A., Pedersen, S., Sorensen, C. B. & Lindelof, P. E. Mesoscopic decoherence in Aharonov-Bohm rings. *Phys. Rev. B* **64**, 045327 (2001).
32  Xiong, J. *et al.* High-field Shubnikov-de Haas oscillations in the topological insulator Bi2Te2Se. *Phys. Rev. B* **86**, 045314 (2012).
33  Sacepe, B. *et al.* Gate-tuned normal and superconducting transport at the surface of a topological insulator. *Nat. Commun.* **2**, 575 (2011).
34  Zhang, Y., Tan, Y.-W., Stormer, H. & Kim, P. Experimental observation of the quantum Hall effect and Berry's phase in graphene. *Nature* **438**, 201 (2005).
35  Novoselov, K. *et al.* Two-dimensional gas of massless Dirac fermions in graphene. *Nature* **438**, 197 (2005).
36  Davies, J. H. *The Physics of Low-dimensional Semiconductors: An Introduction*. (Cambridge University Press 1998).
37  Hikami, S., Larkin, A. I. & Nagaoka, Y. Spin-Orbit Interaction and Magnetoresistance in the Two Dimensional Random System. *Progress of Theoretical Physics* **63**, 707 (1980).
38  Steinberg, H., Laloë, J., Fatemi, V., Moodera, J. & Jarillo-Herrero, P. Electrically tunable surface-to-bulk coherent coupling in topological insulator thin films. *Phys. Rev. B*, 233101 (2011).
39  Lang, M. R. *et al.* Competing Weak Localization and Weak Antilocalization in Ultrathin Topological Insulators. *Nano Lett.* **13**, 48 (2013).
40  Liu, M. H. *et al.* Crossover between Weak Antilocalization and Weak Localization in a Magnetically Doped Topological Insulator. *Phys. Rev. Lett.* **108**, 036805 (2012).
41  Lu, H. Z., Shi, J. R. & Shen, S. Q. Competition between Weak Localization and Antilocalization in Topological Surface States. *Phys. Rev. Lett.* **107**, 076801 (2011).
42  Zhang, L. *et al.* High quality ultrathin $Bi_2Se_3$ films on $CaF_2$ and $CaF_2$/Si by molecular beam epitaxy with a radio frequency cracker cell. *Appl. Phys. Lett.* **101**, 153105 (2012).
43  Qi, X. L., Hughes, T. L. & Zhang, S. C. Topological field theory of time-reversal invariant insulators. *Phys. Rev. B* **78**, 195424 (2008).
44  Schoenberg, D. *Magnetic Oscillations in Metals*. (Cambridge University Press, 1984).